# Comment on "Long-period astronomical forcing of mammal turnover" by van Dam et al., *Nature* 443:687-691

Mensur Omerbashich

*Physics Department, Faculty of Science, University of Sarajevo, Zmaja od Bosne 33, Sarajevo, Bosnia*
*Phone +387-63-817-925, Fax +387-33-649-359, E-mail: momerbasic@pmf.unsa.ba; CC: omerbashich@yahoo.com*

**Summary**

[1] claimed 2.4–2.5 and 1.0 Myr turnover cycles in a Spanish rodent lineages record. However, the record's variance-spectrum, which is missing in [1], shows that the varying reliability and multiple alterations of raw (gapped; unaltered) data by [1] unnaturally boosted the 2.5 and 1.0 Myr noise-cycles to the 99% confidence level, while failing to recognize a third such "99% significant" noise period, of 0.55 Myr at 5.3 var%. Thus at least one claimed period (of 1.0 Myr) is a simple modulation of a relatively stronger noise cycle (of 0.55 Myr) overlooked by [1]. All "99% significant" periods reach mere 5-6% var levels which can be hardly distinguished from noise: those periods' fidelity is at staggering 1-2 orders of magnitude below the usual signal-noise separation marker at 12.0. Remarkably, at least ten noise-periods got boosted to 95% confidence level, and some five noise-periods to near 95% confidence level, as well. Even the zero padding of just 4% of data, as done by [1], significantly suppresses (hence unreported) the strongest 99% significant period, of 7.28 Myr at 7.5 var%. Therefore, the periods claimed are due to strong noise reflection of some intermediary. As hand-waving cyclic-cataclysm claims start to frequent scientific journals, revision of editorial policies is called for on spectral analyses of inherently gapped long records, and of records composed mostly of natural data of significantly inconsistent reliability.

*Keywords*: spectral analysis.

Studies claiming that cataclysmic events are responsible for cyclic variations in long records of natural data appear occasionally in scientific journals. Due to their obvious sensationalism, such reports are almost certain to attract broader public attention. However, long records of natural data in many cases are inherently gapped, and since they are also long the information that they carry are burdened with various influences from many different intermediaries that played a role in the creation of the record.

Unfortunately, it is a common approach in the spectral analysis community to simply proceed to edit such records in order to make them fit the (mostly Fourier) spectral analysis algorithms. This means that the original raw (gapped; unaltered) data and all of data distributions present generally are assumed as entirely understood. One assumption follows another, and it soon becomes easily and erroneously believed by many that data "preparation" could not affect the raw data significantly. Consequently, records end up heavily edited, zero-padded, with values invented, trends subtracted etc.

One such recent study of a mammal record from central Spain [1] reported "new periods" allegedly so close to certain astronomical cycles that a claim of cataclysmic causality was immediately laid. Yet another recent study claimed to have found a new period in a world fossil record [2], which was unlike any other known astronomical cycle but still those authors too made a catastrophic causality claim... In the latter case a closer inspection showed that the cycles claimed by [2] were in fact byproduct of the data treatment applied therein [3]. I show here that the former study too is biased in the same manner, except it produces a result that is coincidentally (by way of a noise intermediary) close to a known astronomical period. The notion of wrong data treatment stands since both of those studies claimed cataclysmic disappearance / reappearance of genera / species, without seriously addressing the characteristics of the spectral analysis technique or its applicability to the data of interest.

I use raw data as the key criterion for assessing a physical claim's validity [3]. Then, Fourier spectral analysis (FSA) is unsuitable for long gapped records [4] such as most of records of natural data. As a result of varying but damaging approaches to data treatment, numerous spectral studies of paleodata range wildly in their conclusions, often going from one extreme ("it's the humans!") to another ("it's Milankovitch!").

To examine how data treatment as done by [1] (averaging to create equispaced values, and alterations to make data fit the FSA equispacing requirement) affected the Spanish record, [1] should have computed that record's variance spectrum; just like the variance is the most natural description of noise in a record of physical data, a variance spectrum tells naturally and simultaneously of the strength of cyclic signals' imprints in noise [3] [5] and thereby of signal's reliability too. I compute variance spectrum of the Spanish record using 99%-confidence strict Gauss-Vaníček spectral analysis [5] [6] [7] [8] [9]. GVSA can analyze raw data without alterations of the input values or output spectra, where spectra can be given in var% [5] [6] or dB [9]. This method has been used over the past thirty years in astronomy, geophysics, biology, medicine, economy and mathematics; it is a spectral analysis method of choice for any science interested in long records of natural data [3]. GVSA is superior to FSA for such records [4], and analogously for records consisted primarily of data of non-uniform reliability, too.

At first look, the variance spectrum of the Spanish rodent lineages record that was used to produce Fig.3b in [1], also seems to be periodic at 99% confidence level, with 2.53 Myr at 6.8 var%, 0.97 Myr at 4.7 var%, and 0.55 Myr at 5.3 var% cycles, see Fig.1 and Table 1. The latter period was not recognized by [1] as being significant, although this period leaves overall a stronger noise imprint than the 1.0 Myr period does, which then almost certainly represents a mere modulation of the overlooked 0.55 Myr cycle. All three periods claimed to be 99% significant reach relatively (compared to 99% confidence level at 4.2% var) weak 5-6% var levels practically indistinguishable from noise, while their fidelity is at insignificant 1-2 orders of magnitude below the usual signal-noise separation marker at 12.0, see Table 1. Data processing, as done in [1] also boosts at least ten spurious periods to 95% confidence level; see Fig.1. In addition, the unresolved portion of the strongest peak is itself seen as reaching almost 95% confidence – together with at least five other noise cycles – indicating that there is no chance that even the strongest claimed period could be real. The primary reason for such a poor performance lies in the interplay of varying reliabilities of thousands of values that were used by [1] to make FSA-fit, i.e., equispaced record of 220 averages. Another reason is in outright data alterations by [1] needed to satisfy operational requirements by the FSA algorithm used. The zero padding is one such alteration commonly performed in Fourier spectral analyses, where inserting zeroes in place of missing values in a time series satisfies the equispacing requirement. But when all nine zero-padded instances are left out from the Spanish record, a forth 99%-significant period, #0, Fig.1, emerges, of 7.28 Myr at 7.5 var%. Meaning, just 9 out of 220 (or 4% of) invented values suppresses the strongest 99%-significant period! Thus, the Spanish record, as a whole, is hugely sensitive to even slightest data manipulations.

Hence, the claimed 99% confidence level is mostly meaningless in the context of total-information quality, in this case. Periods reaching 99% significance indeed leave strong imprints in noise, but this is insufficient for claiming a signal. Hence, the two periods claimed (#1-2, Fig.1) are manifestly noise, of two different kinds: the longer period, #1, represents a robust reflection of some intermediary, and the shorter period, #2, is a mere modulation of its relatively stronger half-period #3. Therefore, no claim in [1] can be true. Treating data as though they were uniformly reliable, as well as performing multiple alterations of raw data, boosts the 2.5 and 1.0 Myr noise-cycles to 99%, but not the 7.28 Myr and 0.55 Myr noise-cycles, which is unexpected as it is a completely arbitrary choice. Note that the latter period, #3, is genuine noise, since, besides having the lowest fidelity of all "99% significant" periods, its fidelity is also at negligible 0.007. Having the absolutely largest fidelity, the period #0 could be the most real one of all four 99% significant periods; however this statement should not be taken literally either, since #0 too was probably caused by some data alterations other than the zero padding which suppressed it.

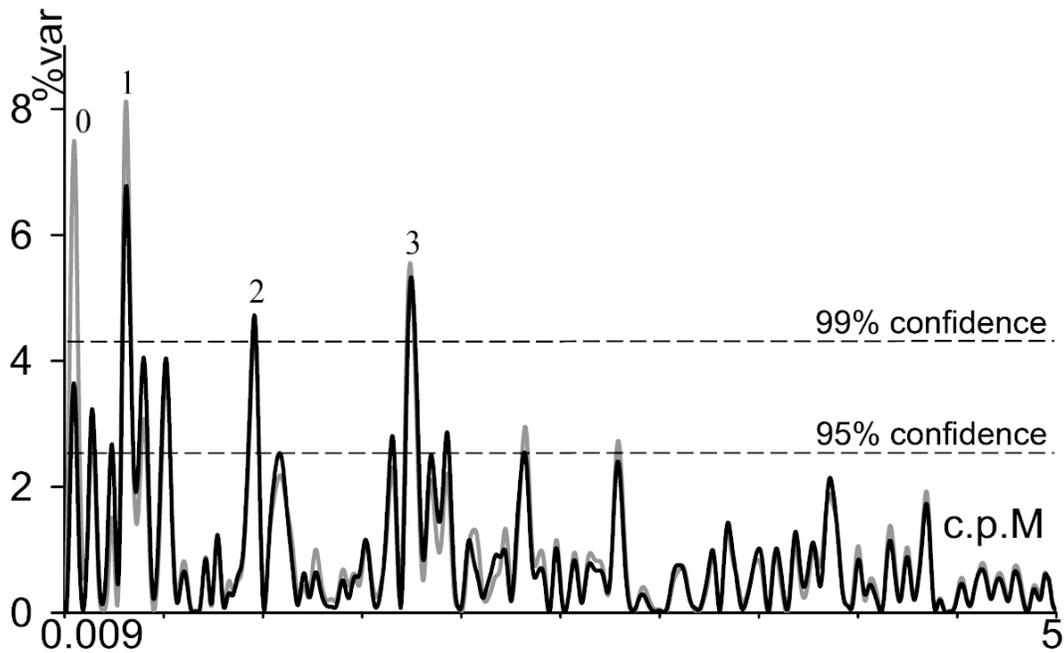

Figure 1. Variance spectrum of the Spanish record after it was altered by [1] (black), v. variance spectrum of the same record excluding zero-padded "values" (gray). Gray line shows the effect of zero padding on a long gapped record of natural data. Frequencies in cycles per 1 Myr (c.p.M). Periods #0-3 listed in Table 1. Spectral resolution for plotting: 2000 spectral points.

| spectral peak # | period [Myr] | fidelity | mag [var%] | mag [dB] |
| --- | --- | --- | --- | --- |
| 0 | 7.28 | 1.211 | 7.50 | -10.91 |
| 1 | 2.53 | 0.146 | 6.78 | -11.38 |
| 2 | 0.97 | 0.022 | 4.73 | -13.04 |
| 3 | 0.55 | 0.007 | 5.33 | -12.50 |

Table 1. Values corresponding to three 99% significant periods, Fig.1. Spectral resolution for computations: 20 000 spectral points.

## Conclusion

Experience teaches us that even a single catastrophic event claim, when promoted in the media, could result in havoc. It seems inevitable that researchers could soon start making "predictions" out of the many reports alleging some "99% certain" past recurring cataclysms. As the hand-waving cyclic-cataclysm reports start to frequent scientific journals, revision of editorial policies is called for in cases of the spectral analyses of long and inherently gapped records of natural data, or, more generally, records that contain natural data most of which have significantly varying reliability. Such revisions could entail approaches presently unthinkable of when refereeing those reports, like the imposing of mandatory blind test(s) using synthetic data, or/and repetitions of critical computations, or/and testing results (using independent methods) for spectrum distortion due to data manipulations, and so on. Although seemingly elementary, such measures could prevent researchers from missing a bigger picture, as they get technical.

Spectral analysis is at least as much art as it is science, requiring various choices to be made prior to punching the data into an algorithm. Proper choice of the analysis technique (algorithm) must not be the first but instead the last of those choices to be made. This should be preceded by considerations such as selection of the criteria for data treatment approaches to be used. (A fundamental such criterion, of using raw data, was applied here.) Hopefully, this could prevent blunders like [1] from ever entering into press. The primary goal of scientific analyses of physical time series should be to responsibly produce publicly useful information on cyclic natural phenomena, so that science is not being undermined by the ill reputes of cyclic failures. This call for action coincides in time with challenges faced by humankind in taking full responsibility (instead of accusing nature and divine acts alike) for harming life and environment on Earth.